\documentclass[aip,reprint]{revtex4-1}
\usepackage{graphicx}
\usepackage{dcolumn}
\usepackage{bm}
\usepackage{comment}  
\usepackage[utf8]{inputenc}
\usepackage[T1]{fontenc} 
\usepackage{mathptmx}
\usepackage{etoolbox}
\usepackage{color}
\draft 
\newcommand{\mmax}[1]{{{\color{black}#1}}}

\begin{document}

\title{Haptic Light-Emitting Diodes: Miniature, Luminous Tactile Actuators}

\author{Max Linnander}
\author{Yon Visell}
\altaffiliation[Also at ]{Department of Bioengineering, Media Arts \& Technology Program, and Department of Electrical and Computer Engineering, University of California, Santa Barbara, USA.}
\affiliation{Department of Mechanical Engineering, University of California, Santa Barbara, USA}


\date{\today}

\begin{abstract}
We present Haptic Light-Emitting Diodes (HLEDs), luminous thermopneumatic actuators that directly convert pulsed light into mechanical forces and displacements.  Each device packages a miniature surface-mount LED in a gas-filled cavity that contains a low-inertia graphite photoabsorber.  The cavity is sealed by an elastic membrane, which functions as a working diaphragm.  Brief optical pulses heat the photoabsorber, which heats the gas.  The resulting rapid pressure increases generate forces and displacements at the working diaphragm.  Millimeter-scale HLEDs produce forces exceeding 0.4 N and displacements of 0.9 mm at low voltages, with 5 to 100 ms response times, making them attractive as actuators providing tactile feedback in human-machine interfaces.  Unusually, these actuators are also light-emitting, as a fraction of optical energy is transmitted through the membrane.  These photomechanical actuators have many potential applications in tactile displays, human interface engineering, wearable computing, and other areas.
\end{abstract}


\maketitle 

Converting optical energy directly into mechanical work in compact, manufacturable devices would bridge solid-state optoelectronics with mesoscale actuation regimes, including displacements and forces beyond the reach of micro-electromechanical (MEMS) actuators.  Haptics provides a stringent and well-characterized target for the design of such transducers. 
Practical, high-fidelity tactile actuation demands localized actuation via interfaces on the millimeter-scale that operate at moderate voltages and can generate substantial out of plane displacement ($10^{-3}$ m) and forces ($10^{-2}$ to $10^{-1}$ N), with short activation response times ($10^{-3}$ to $10^{-1}$ s).\cite{jones2006human,copeland2010identification,craig1987vibrotactile}    

Existing actuator families meet subsets of these requirements but involve familiar tradeoffs. Electromagnetic devices serve most needs in mechatronics, but confront current-density and thermal limits under miniaturization.\cite{kastor2023ferrofluidic,streque2012emactuator} Electrostatic and dielectric-elastomer actuators offer high bandwidth and strain but typically require kV-scale drive voltages, complicating insulation and packaging.\cite{leroy2020haxel,grasso2023haxel,mitchell2022HV} Piezoelectric actuators provide fast transduction but generate very small displacements, and introduce significant mechanical and electronic constraints.\cite{wood2005piezeo} Emerging approaches, including electroactive polymers and electroosmotics, show promise but rely on specialized materials and fabrication processes.\cite{biswas2019materials,schultz2023flatpanel}

\begin{figure*}
    \includegraphics[width=\textwidth]{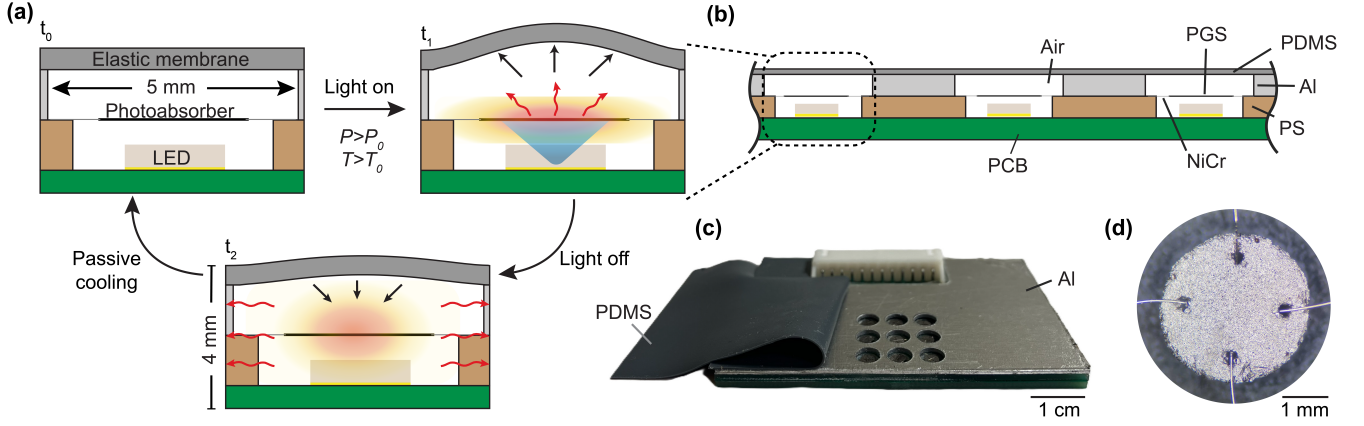}
    \caption{Design and operating principle of the Haptic Light-Emitting Diode. (a) Light from the LED is converted into heat by the photoabsorber.  Heat is transferred to air in the cavity, raising gas temperature $T$ and pressure $P$.  Gas expansion drives the deflection of the elastic membrane. (b) The HLED is an assembly of patterned layers, supporting manufacturability. From top: Elastic membrane (PDMS), cavity walls (Aluminum), thin photoabsorber (pyrolytic graphite sheet suspended on nichrome wires), cavity walls (PS), PCB with an LED. (c) Photo of a 3x3 array of HLEDs with the elastic membrane partially removed. (d) \mmax{Suspending the PGS photoabsorber on NiCr wires favors heat transfer to the air.}}
    \label{Fig:concept}
\end{figure*}

Photo-thermomechanical actuation methods -- based on optically driven thermoelastic expansion, thermal bilayers, or thermal gas expansion -- have been investigated for microvalves, Braille displays, and soft robotics, among other uses.\cite{mckenzie1995design,torras2014optomechanical,paul2024photoactuator,camargo2012opticalbraille,hiraki2020laserpouch,hwang2021bimorph} Many reported mechanisms deposit the majority of heat in liquid or solid materials, whose dynamics are governed by the thermal time scale, $\tau$, that scales with thermal mass, $C$. The large heat capacities of solids and liquids ($\sim$1 J/cm$^3$K) and strong thermal coupling to supporting structures slow transients and shunt energy before it can perform useful work, yielding response times of $10^0-10^2$ s.\cite{camargo2012opticalbraille,hiraki2020laserpouch} Thermopneumatic approaches offer the potential for faster responses by using a captive working gas to perform mechanical work. Although such configurations have been studied, the heating element is often integrated in the wall of the gas-filled cavity, yielding long response times and low forces.\cite{mazzotta2023thermopneumatic,mazzotta2025thermopneumatic}

Such issues could be overcome by directly heating a captive working gas via light absorption.  Unfortunately, gases are insufficiently opaque outside of exotic conditions.  An approximation to this ideal might involve radiative heat transfer to a low-thermal-inertia, large-aspect-ratio photoabsorber suspended within the gas. Such a design would provide for rapid transfer of optical energy to the gas. This principle was recently exploited for remotely actuating mesoscale architected surfaces, but required external optoelectronic and optomechanical components.\cite{linnander2025tactile}

Here, we present Haptic Light-Emitting Diodes (HLEDs) -- millimeter-scale luminous thermopneumatic actuators that encapsulate miniature surface-mount device (SMD) power LEDs in packaging that facilitates photomechanical energy conversion.  Within each pixel, an LED emits pulsed light on application of a brief current pulse, 5 to 100 ms, supplied at low voltage (4 V).  The majority of emitted light is absorbed by a graphite film photoabsorber suspended within the cavity. The photoabsorber has minimal thermal inertia and is rapidly heated, quickly transferring heat to the small volume of encapsulated gas. Gas temperature and pressure increase over a timescale of tens of milliseconds, driving gas expansion and mechanical deflection of the membrane sealing the cavity, which functions as the working diaphragm of the actuator.  \mmax{As described below, the photoabsorber is suspended by nichrome wire (AWG 48 in our implementation) with orders of magnitude smaller cross section area than the photoabsorber.  This ensures that the majority of heat added to the photoabsorber is transferred to the gas, rather than along the wires, enabling} light pulses with energies of 1 to 200~mJ to generate substantial forces up to 440 mN, and displacements of nearly 1 millimeter.  The fast response time of the actuator and high rates at which it can be driven (up to 200~Hz) meet requirements for high-fidelity tactile feedback, and hold promise for applications in other domains.  The pixels emit a fraction of emitted light scattered in the cavity, which lends them unique capabilities for simultaneous optical and mechanical feedback. The same techniques used here could be scaled to much larger arrays using standard techniques. 

Figure \ref{Fig:concept}(a) illustrates the design, operating principle, and physical process. Each HLED acts as an episodically driven thermodynamic engine driven by the absorption of pulsed light.  Due to the low net heat capacity of the photoabsorber, its temperature rapidly increases by several hundred degrees Celsius under illumination. Heat is rapidly transferred to the gas, transiently increasing gas pressure. Because the cavity dimensions are small, the characteristic (acoustic) time for pressure to equilibrate is small, $\tau_\mathrm{eq} \ll 1$ ms, relative to the time scales of interest, thus pressure is effectively constant across the cavity. 
Gas expansion drives expansion of the working membrane, yielding forces $F(t)$ and displacements $z(t)$. At the end of the light pulse, heat in the air is transferred through the cavity walls to the cooler environment, restoring the device to its initial state over a fraction of a second.

This general strategy may be used to realize a large variety of actuators. The following design choices exemplify one way HLEDs can be implemented. A multi-layered architecture was used to facilitate ease of manufacturing and scalability (Fig. \ref{Fig:concept}(b-c)). A printed circuit board (PCB, thickness: 1.6~mm) with an SMD LED forms a base layer. An array of air-filled cylindrical cavities is formed from one layer of polysiloxane (PS, diameter:~4 mm) and one layer of aluminum (Al, diameter: 5~mm) that are patterned via laser cutting and drilling. Between the pixel cavity layers is a pyrolytic graphite sheet (PGS, 17~$\mu$m) that is cut in a circular shape (diameter: 3.3~mm) using a vinyl cutter and suspended on 40~AWG Nichrome (NiCr) wires (Fig. \ref{Fig:concept}(d)). PGS was selected due to its high thermal stability and broad spectrum absorption \cite{linnander2025tactile}, and NiCr for its oxidation resistance. \mmax{Suspending the PGS on thin NiCr wires was done to favor heat transfer to the air.} Capping this assembly is the elastic membrane -- a 250~$\mu$m polydimethylsiloxane (PDMS) layer that seals the air-filled cavity. \mmax{The full assembly has thickness 4~mm.  The planar dimensions of the 9-pixel assembled prototype are 5 cm $\times$ 4 cm (Fig. \ref{Fig:concept}(c)). Each pixel aperture has diameter  5~mm.} The display can be made flexible through the use of compliant materials. Additional fabrication details are in the Supplemental Materials. 

HLEDs are stimulated by brief light pulses (0.5 to 100~ms) delivered from a blue LED ($\lambda=$ 458~nm, Cree XEG), which is driven by a MOSFET. The control signal is delivered from a data acquisition device (National Instruments USB DAQ) using a pulse-width modulation signal at the MOSFET gate (Fig. S2). Concurrently, localized visible light is emitted due to the transmission of scattered light through the top membrane. \mmax{The intensity of the transmitted light can be tuned by engineering the opacity of the top membrane.}

\begin{figure}
    \includegraphics[width=8.5cm]{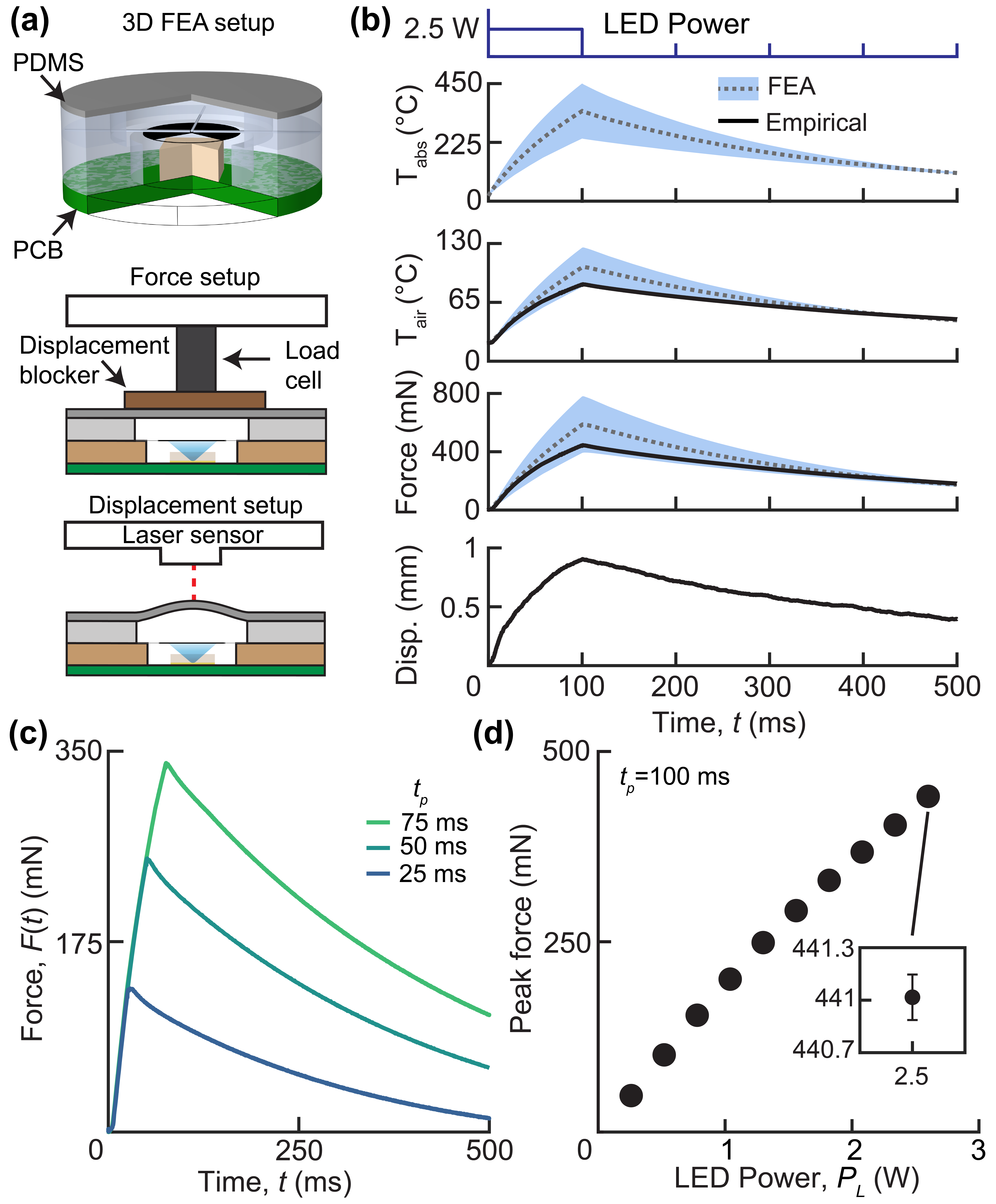}
    \caption{
    (a) \mmax{Illustrations of the }experimental configurations: Finite element analysis (FEA), isometric force measurement, and free-displacement measurement. (b) Photoabsorber temperature $T_\mathrm{abs}(t)$, cavity air temperature $T_\mathrm{air}(t)$, isometric force $F(t)$, and free displacement $z(t)$, for a 100-ms, 2.5-W optical pulse. Solid black lines are laboratory measurements, except for air temperature, which is obtained via the ideal gas law. Numerical (FEA) results: Dashed line - mean response; Shaded region - parametric sweep over parameter uncertainties, see Table S1 \& S2. (c) Force, $F(t)$, for various pulse durations at optical power $P_L=2.5$~W. Laboratory measurements. (d) Peak measured force as a function of LED optical power $P_L$ (Error bar: standard deviation, $n=4$).}
    \label{Fig:mechanical}
\end{figure}

We characterized the thermal and mechanical responses using numerical finite element analysis (FEA) and laboratory measurements under constant power pulse driving (pulse duration  $t_p$, power $P_L$; Fig. \ref{Fig:mechanical}(a)). In the simulations, geometric and material properties were swept across a range reflecting manufacturing and geometric uncertainties, yielding an envelope of thermomechanical response behaviors (Table S2). In laboratory testing, force was isometrically measured using a load cell (Futek LSB200).  Unloaded displacement was measured at the center of the membrane aperture using a laser displacement sensor (Keyence IL-065). In both simulations and lab experiments, air temperature, $T_\mathrm{air}(t)$, was computed from measured force using an analytical heat transfer expression derived from the ideal gas law, $T_\mathrm{air}(t)=T_0\Big(F(t)/(\pi(d/2)^2\cdot P_0)+1\Big)$, where $d$ is the cavity diameter at the interface with the elastic membrane, and $P_0$ and $T_0$ are the initial pressure and temperature in the cavity. 

We observed the mean temperature, $T_\mathrm{abs}(t)$, in the photoabsorber to rise monotonically during light exposure, driving monotonic increases in cavity air temperature and pressure.  The resulting force, $F(t),$ at the working membrane yielded unloaded displacement, $z(t)$ (Fig. \ref{Fig:mechanical}(b)). Temperature, force, and displacement reached maximum values near the time light exposure ceased. The relationship between photoabsorber temperature, pressure, and force was thus nearly static, as has been observed in thermopneumatic devices with similar dimensions.\cite{linnander2025tactile} Heat transfer from the suspended photoabsorber to the environment can be modeled via an effective thermal circuit. The solution for photoabsorber temperature has the form
\begin{equation}
    T_\mathrm{abs}(t) = \varepsilon P_LR\Big(1-\exp(-t/\tau)\Big)+T_0\exp(-t/\tau),
    \label{eq:T_abs}
\end{equation}
Here, $\varepsilon$ is the fraction of incident power absorbed; for PGS at $\lambda = 458$~nm, $\varepsilon = 0.72$.\cite{linnander2025tactile}  $R$ is the effective thermal resistance determined by heat transfer to the air and through the NiCr wires. The thermal time constant, $\tau=RC$, where $C$ is the heat capacity, is the time scale of relaxation of the photoabsorber temperature toward the ambient temperature, $T_0$. In contrast, heat can be injected at any rate (subject to constraints) via light absorption. We computed emitted optical power, $P_L$, for each experiment using the LED manufacturer specifications, referenced to the applied driving current.\cite{cree_xlamp_xeg} The maximum instantaneous current used in this research was $I_\mathrm{max} = 2400$~mA, corresponding to emitted optical power $P_L \approx 2.5$~W.  The amount of this power converted into heat in the photoabsorber was 1.8~W.

Experimental results were consistent with simulations and qualitative predictions from theory.  At optical power  $P_L = 2.5$~W and pulse duration  $t_p = 100$~ms, numerical experiments yielded photoabsorber temperatures of $340~^\circ$C. 
The corresponding laboratory experiments yielded a peak air temperature of 85~$^\circ$C (computed from the isometric force), a peak force of 440~mN, and a peak displacement of 0.9~mm. After the LED was switched off, the system returned to its original state with a cooling time constant of $\tau = 440$~ms (determined by fitting eq. \ref{eq:T_abs} to the data, $r^2=0.99$). With power held constant, the peak force increased with pulse duration (Fig. \ref{Fig:mechanical}(c)). Similarly, with pulse duration held constant, we observed the peak output force to increase monotonically with increasing optical power (Fig. \ref{Fig:mechanical}(d)), as expected from Eq. 1. Mechanical output of the HLED was highly consistent; consecutive pulses at $P_L = 2.5$ W varied in force output by less than 0.1\%. 

\begin{figure}[b]
    \includegraphics[width=8.5cm]{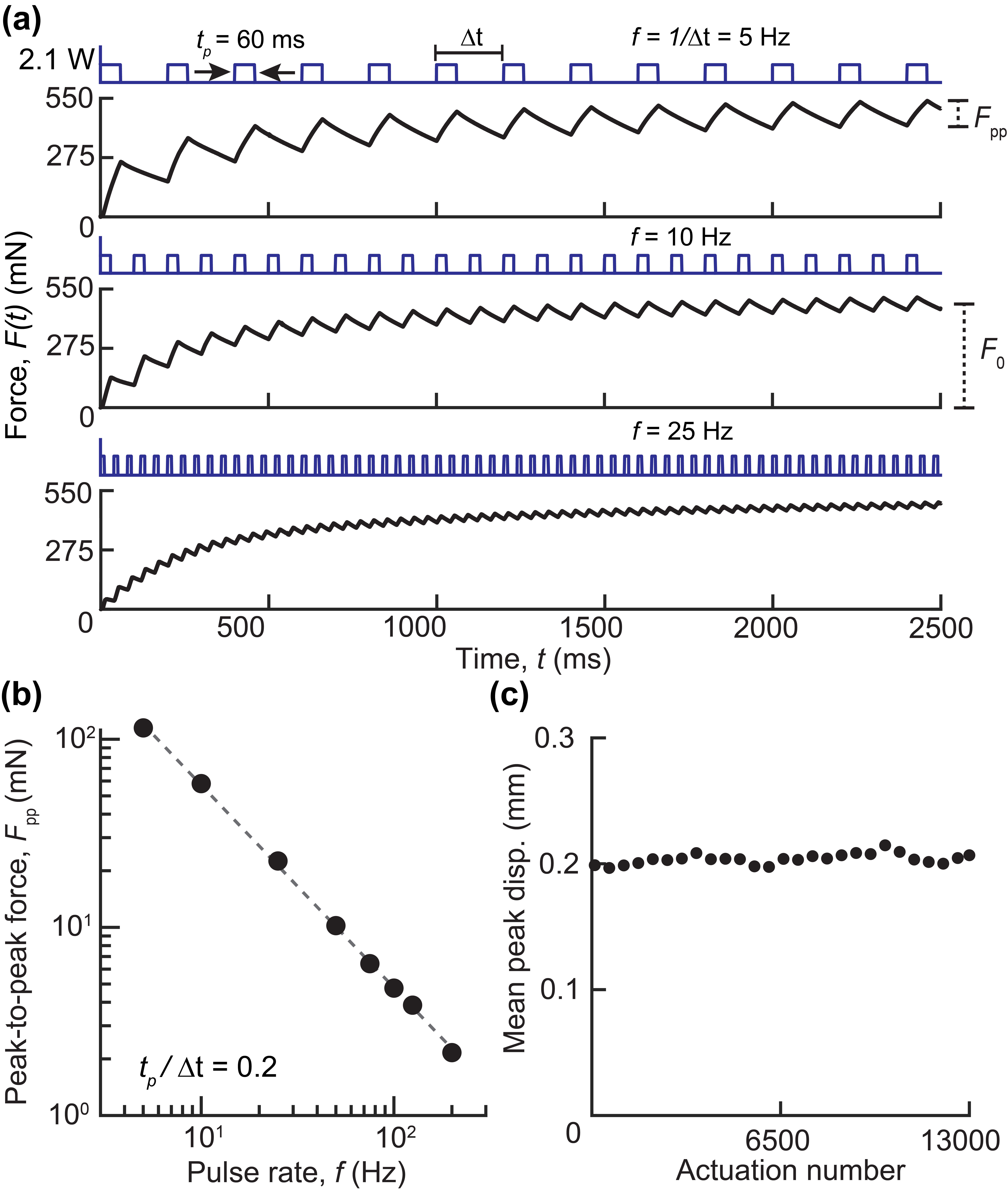}
    \caption{
    (a) Isometric force responses to pulse trains at various pulse rates, $f$, with duty cycle $t_p/\Delta t=0.3$. (b) Peak-to-peak force component $F_{pp}$ as a function of pulse rate $f$ with duty cycle $t_p/\Delta t=0.2$. Regression fit: $\log_{10}(F_{pp})=\alpha\log_{10}(f)+\beta$, $\alpha=-1.08$, $\beta=2.83$, $r^2=0.99$. (c) Mean free displacement is consistent over 13,000 actuations. Each dot corresponds to the mean of 100 peaks.}
    \label{Fig:dynamic}
\end{figure}

Response time and operating frequency range are key figures of merit for tactile devices.  During manual activities, the sense of touch exploits temporal content with millisecond resolution in gauging the texture of surfaces, detecting the onset of skin-object contact in grasping, and in many other manual tasks.  Touch is thus sensitive to wide-band frequency content up to several hundred Hz. Consequently, actuator dynamics is paramount in the engineering of tactile devices.  

As seen in Fig. \ref{Fig:dynamic}, HLEDs exhibit rapid response times of 5 to 100 ms, and a substantial operating frequency range, up to 200 Hz.  The characteristic time scale for thermal cooling can be tuned, but in the reported devices is approximately $\tau \approx 440$ ms, due to the large thermal resistance between the photoabsorber and the surrounding materials, which are separated by an air gap. If two impulses are produced in succession with inter-pulse time  $\Delta t < \tau$, residual heat from the first impulse biases the operating point of the second.  This leads to a bimodal response comprising a slowly varying component, $F_0$, and a rapidly oscillating component that tracks the pulse rate, $F_{pp}$;  similar phenomena were observed in various systems.\cite{hwang2021bimorph,mazzotta2025thermopneumatic}. \mmax{A prior study involving a similar configuration, demonstrated that this process is indeed explained by fundamental heat transfer, and that relative proportion of slow and oscillating responses can be adjusted by varying the ratio, $(\Delta t-t_p)/\tau$, of the time in between pulses, $\Delta t -t_p$, to the timescale $\tau$ for heat transfer from photoabsorber to cavity walls, $\tau = R C$.\cite{linnander2025tactile}} Regardless, pulsing at rates faster than $1/\tau$ is appropriate for tactile applications, since the sense of touch is far more sensitive to forces oscillating at frequencies of hundreds of Hz than it is to slow (near DC) frequencies.\cite{morioka2005thresholds} 

We thus measured the rapidly oscillating component, $F_{pp}$ for pulse rates, $f$, between 5 and 200~Hz (Fig. \ref{Fig:dynamic}(b)). Across this range, the peak-to-peak amplitudes for force ranged from 113~mN down to 2.2~mN, whereas the displacement ranged from 170~$\mu$m down to 4.4~$\mu$m (Fig. S1). 
Both force and displacement amplitudes scale approximately as $f^{-1}$, corresponding to a slope of $-1$ in log--log space, consistent with a scaling relation $E \propto 1/f$ of per-pulse energy.

We actuated the HLED for 13,000 cycles, using low-duty-cycle pulses, and observed no change in displacement amplitude or material degradation (Fig. \ref{Fig:dynamic}(c)). This outcome is consistent with material limits: in air, graphite oxidation mainly occurs at sustained temperatures above 400~$^\circ$C.\cite{xiaowei2004oxidation} The photoabsorber likely maintained a temperature well below this threshold, as indicated by Fig. \ref{Fig:mechanical}(b). Likewise, the mean cavity air temperature reached a peak of 84~$^\circ$C, so we expected no degradation of the adhesives or cavity walls, which had temperature ratings of 200~$^\circ$C.

\begin{figure}[t]
    \includegraphics[width=8.5cm]{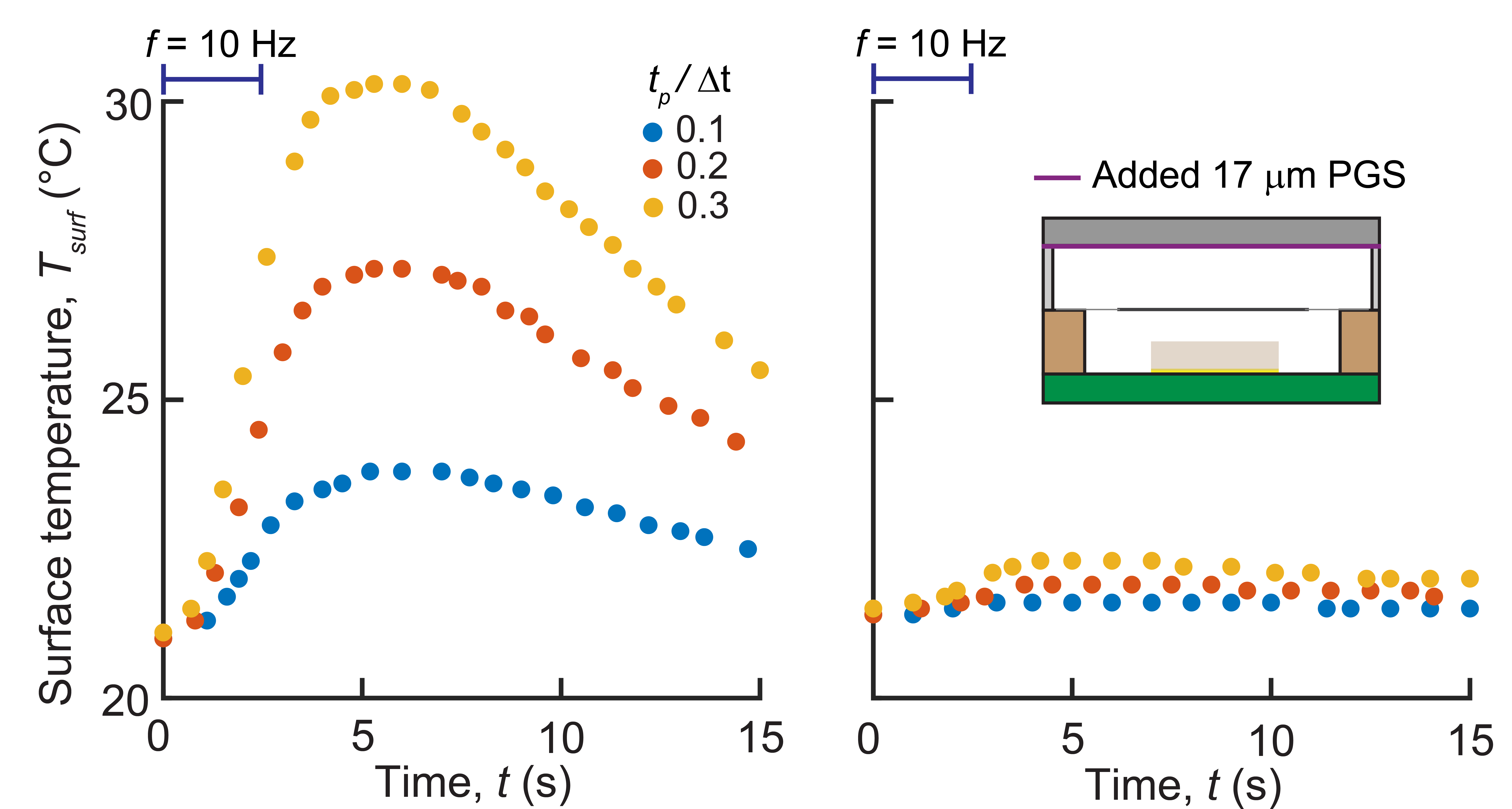}
    \caption{(a) Transient surface temperature rise for various duty cycles, $t_p/\Delta t$, during a 2.5-s actuation window, followed by passive cooling. (b) Adding a PGS layer beneath the PDMS membrane suppresses the temperature rise across all duty cycles.}
    \label{Fig:thermal}
\end{figure}

In prototyping, we observed that continuous pulsing resulted in elevated surface temperatures (Fig~\ref{Fig:thermal}(a)).  Thus, we integrated a layer of PGS, 17 $\mu$m in thickness, immediately below the elastic top membrane (Fig.~\ref{Fig:thermal}(b)), exploiting its high in-plane thermal conductivity. As shown in the figure, prior to adding the PGS layer, the surface temperature increase was $\Delta T_\mathrm{surf} = 9.2$~$^\circ$C after 2.5 seconds of operation.  With the PGS layer integrated, the temperature increase was reduced to $\Delta T_\mathrm{surf} = 0.8$~$^\circ$C, which is comparable to the amount of heating that would be supplied by a human finger in contact with the membrane \cite{linnander2025tactile}.  The effect on blocked force was negligible (Fig. S3(a)), even though we took no measures (e.g. perforation) to ensure that the PGS sheet was gas permeable. Pulse driving the improved HLED continuously over a duration four times longer, for 10 seconds, yielded a temperature increase of $\Delta T_\mathrm{surf} = 1.5$~$^\circ$C that is very slight, particularly for an electronic device such as an actuator, but is perceptually noticeable (Fig. S3(b)).  To place these observations in context, we note that it is rarely desirable for tactile actuators to be pulsed continuously for more than a fraction of a second, due to the annoyance it causes and due to sensory adaptation effects.\cite{verrillo1977effect} Indeed, modern consumer devices rarely output sustained buzzing, but rather use very brief vibrational pulses for notifications. \mmax{If pulsing for several seconds or more were required, in other applications that might require the exterior surface temperature of the device to be sufficiently low, enhanced passive or active cooling could be applied to the housing, at the expense of further device complexity.  }

\begin{figure}[b]
    \centering
    \includegraphics[width=0.85\linewidth]{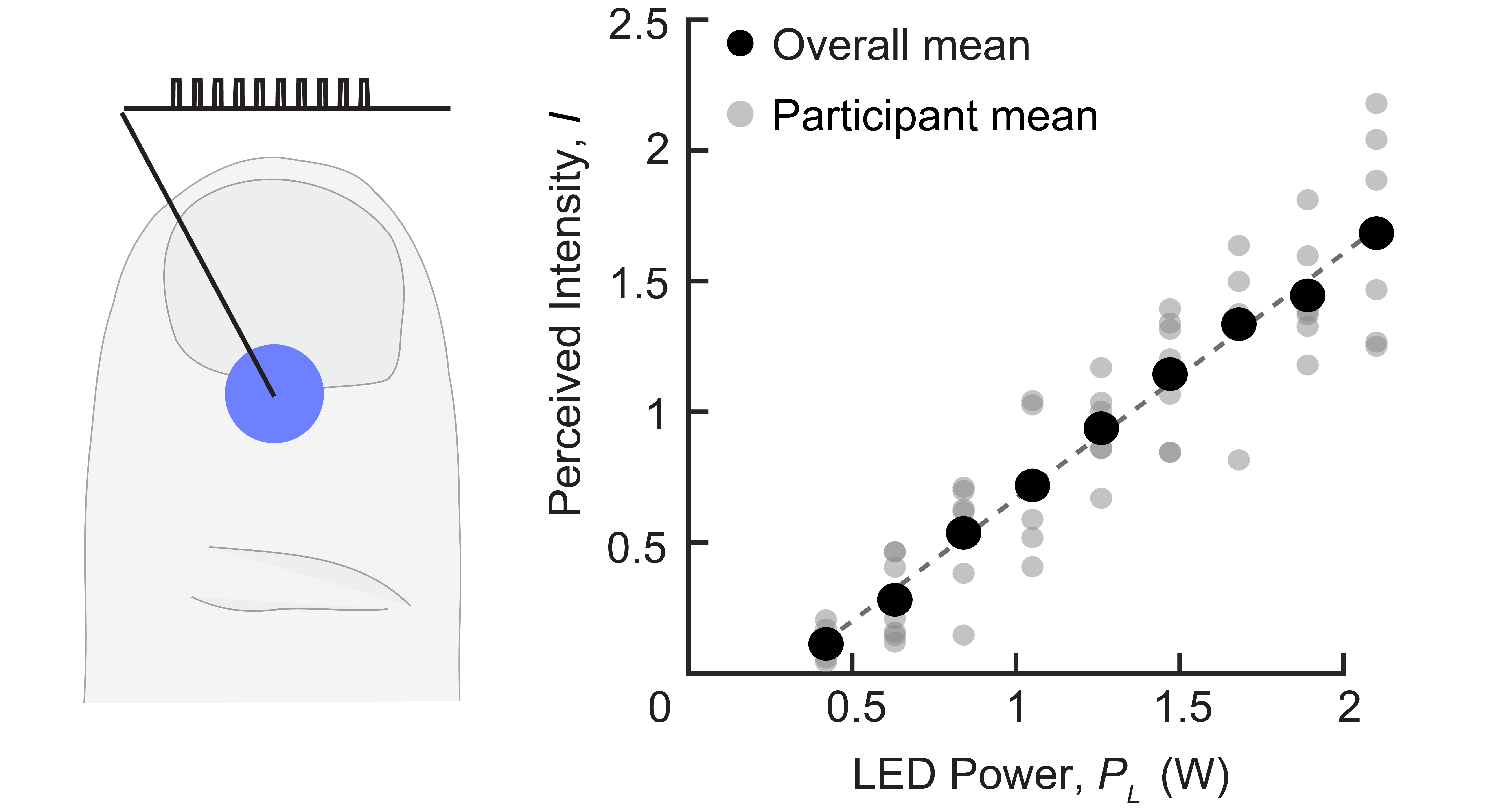}
    \caption{Perceived intensity as a function of power, measured using the method of magnitude scaling.\cite{jones2013psychophysics,stevens1958problems}.
    Regression fit: $I = \alpha P_L+\beta$, $\alpha = 0.0197$, $\beta = -0.2693$, $r^2 = 0.99$.}
    \label{fig:perceptual}
\end{figure}

The forces and displacements the HLED generates are well within the perceptual range \cite{jones2006human}. We experimentally measured perceived tactile intensity $I$ as a function of optical power $P_L$ in a perception experiment, finding that perceived intensity, $I(P_L)$, increased linearly with power (Fig.~\ref{fig:perceptual}, Supplementary Materials).  Using the results, a specified level $I$ of perceived strength can be attained by driving the HLED with the corresponding power $P_L(I)$. The experiment used the psychophysical method of magnitude estimation, a well-established quantitative technique in perception research.\cite{jones2013psychophysics,stevens1958problems} 
 
The substantial range of forces and short response times of HLEDs compare favorably with \mmax{prior} tactile actuators at this scale \mmax{(Table S4)}. \mmax{HLEDs} yield forces that are an order of magnitude larger \mmax{than comparable  thermopneumatic devices from prior studies.}\cite{linnander2025tactile,mazzotta2023thermopneumatic,mazzotta2025thermopneumatic} While HLEDs were designed for haptic applications, the same general design strategy could be \mmax{used to realize devices for use} in other domains.  By increasing size, and altering the geometry and configuration, \mmax{the range of forces and displacements can be increased}, which may make these devices suitable for applications in microrobotics and microfluidics, among others.  Indeed, commercial LEDs of similar size to those used here provide optical power more than 10$\times$ higher than is used here. 

In conclusion, we introduced compact thermopneumatic actuators, Haptic Light-Emitting Diodes (HLED).  These devices consist of packaging supplementing miniature SMD power LEDs, converting optical energy into mechanical work through thermodynamic gas expansion.
The devices operate at 4 V, exhibit millisecond response times, displacements approaching 1 mm, forces up to 440 mN, and support actuation rates from 1–200 Hz. A unique advantage of these actuators is that they also output light, similar to a standard LED, thus furnishing visual and mechanical output. These findings highlight the broader potential of optically driven thermopneumatic actuation for compact, high-performance devices across a range of emerging applications.

\subsection*{Supplementary Material} 
\mmax{The supplementary material provides additional detail on the numerical methods, perceptual testing, photomechanical energy conversion, and potential future research directions.}

\subsection*{Author Declarations}
\subsubsection*{Conflict of Interest}
The authors have no conflicts to disclose.

\subsubsection*{Author Contributions}
\textbf{Max Linnander:} Conceptualization (equal); Data curation (equal); Investigation (equal); Writing original draft (equal); Writing–review\&editing (equal). \textbf{Yon Visell:} Conceptualization (equal); Supervision (lead); Validation (equal); Writing original draft (equal); Writing–review\&editing (equal).
\subsubsection*{Data and materials availability:}
All data needed to evaluate the conclusions have been made publicly available on Zenodo: 10.5281/zenodo.18222861.

\bibliography{refs}

\end{document}